\documentclass[fleqn,10pt]{wlscirep}
\usepackage[utf8]{inputenc}
\usepackage[T1]{fontenc}
\usepackage{hyperref}
\usepackage{algorithm}      
\usepackage{algorithmic}

\title{The impact of anticonformity on the diffusion of innovation -- insights from the $\mathbf{q}$-voter model}

\author[1,*]{Angelika Abramiuk-Szurlej}

\affil[1]{Department of Operations Research
and Business Intelligence, Wroc\l{}aw University of Science and Technology, 50-370 Wroc\l{}aw, Poland}

\affil[*]{angelika.abramiuk-szurlej@pwr.edu.pl}

\keywords{Collective adaptation, Diffusion of innovation, Agent-based model, Social network, Mean-field approximation}

\begin{abstract}
Anticonformity, behaving in deliberate opposition to the group of influence, has long been recognized as a distinct social response, differing both from conformity and from independence. While often treated as a source of noise or contrarianism, anticonformity can play a constructive role in social dynamics by counterbalancing majority pressure and influencing collective outcomes. Recently, it was shown in laboratory experiments that evaluation may induce strategic anticonformity when rewards are anticipated. Moreover, using agent-based modeling, it has been demonstrated that anticonformity can depolarize highly polarized social groups and prevent social hysteresis. These findings encouraged us to extend the $q$-voter model with asymmetric independence, an agent-based model of the diffusion of innovation, by introducing anticonformity, so that agents can act independently, follow the group, or oppose it. Using a mean-field approximation (MFA), we investigate how these behavioral tendencies influence the diffusion of innovation. Our results show that anticonformists can accelerate early adoption and enable successful diffusion even in cases where diffusion would otherwise fail. The model exhibits two stable adoption levels separated by an unstable branch, giving rise to hysteresis and a critical mass effect. We also demonstrate that increasing independence lowers the threshold of anticonformity needed for widespread adoption. These results highlight how anticonformist behavior can facilitate innovation diffusion, with practical implications for decision-makers and policy design.
\end{abstract}

\begin{document}

\flushbottom
\maketitle

\thispagestyle{empty}

\section{Introduction}
\label{sec:Introduction}
Anticonformity, the deliberate choice to behave in direct opposition to a majority or peer group opinion, is a fundamental social response that has long been distinguished from both conformity and independence in social psychology and complex systems research \cite{Nail20131,Starnini2025}. While conformity involves alignment with the majority, and independence is behavior based solely on internal judgment, anticonformity is a reactive stance, a repulsive social influence that actively resists social pressure. In opinion dynamics models this behavior was often dismissed as mere contrarianism or noise in social dynamics \cite{Gal:25,Gal:24,Gal:20,Nyczka2013,Jed:Szn:19,Starnini2025}.

However, a growing body of work underscores the constructive and critical role anticonformity can play in shaping collective outcomes. By counterbalancing strong majority pressure, anticonformity can prevent social traps and rigid consensus formation. Recent laboratory experiments have shown that individuals can engage in strategic anticonformity when placed in situations of evaluation, deliberately standing out to attract anticipated rewards or avoid punishment \cite{Dvorak2025}. Furthermore, in computational models of opinion dynamics, this phenomenon has been demonstrated to act as a powerful depolarizing force in highly polarized social groups \cite{Lip:Szn:25} and can also prevent the emergence of social hysteresis, a phenomenon where a system's state lags behind changes in the external factors that caused it \cite{Kaminska2025}.

Despite its recognized importance in shaping opinions, the potential role of anticonformist behavior in the context of innovation diffusion remains largely overlooked, especially in computational models. These findings encouraged us to ask how anticonformity influences the diffusion of innovation, which involves the spread of new ideas, technologies, or practices through a population and critically depends on how individuals respond to the choices of others \cite{Rog:03,Kuandykov2010531,Rand2021}.

To address the question of the role of anticonformity in innovation diffusion, we use the $q$-voter model with asymmetric independence \cite{abr-szu:wer:arxiv:25} and extend it by introducing anticonformity. Using a mean-field approximation (MFA), we analyze the model and show that anticonformity can accelerate early adoption and, critically, enable successful widespread diffusion even in scenarios where the innovation would otherwise fail to reach a critical mass. We also show that the model exhibits two stable adoption levels separated by an unstable branch, giving rise to hysteresis and a critical mass effect, and that the interplay between independence and anticonformity is key to determining the ultimate success of the innovation. These results provide new insights into the dynamics of collective behavior and offer practical implications for decision-makers and policy design focused on facilitating the adoption of new technologies and ideas.

\section{The model}
\label{sec:Model}
We consider a population of $N$ agents distributed at the nodes of an undirected network. The connections between agents represent the structure of social relations within a given social system. Every agent $x$ is characterized by a binary state variable $S_x = \pm 1$, where $S_x = +1$ denotes adoption of an innovation, and $S_x = -1$ represents non-adoption. The system evolves in discrete time steps according to three basic mechanisms: independent behavior and social influence, in the form of conformity or anticonformity.

At each time step, we randomly select one agent $x$ to potentially revise its state. With probability $p^{\mathrm{ind}}$, the agent acts independently of its neighbors. In this case, it adopts the innovation with probability $p^{\mathrm{eng}}$ or becomes non-adopted with probability $1 - p^{\mathrm{eng}}$. With complementary probability $1 - p^{\mathrm{ind}}$, the agent is subject to social influence. A group of $q$ neighbors is randomly selected without replacement from agents directly connected to the agent. If all $q$ neighbors share the same opinion, the agent adopts that opinion with probability $p^{\mathrm{con}}$ or acts oppositely to the group with probability $p^{\mathrm{ant}}$; otherwise, it retains its current state. Of course, there should be $p^{\mathrm{ind}} + p^{\mathrm{con}} + p^{\mathrm{ant}} = 1$, so that at each update the agent’s behavior is determined by one of the three mutually exclusive mechanisms: independent action, conformity, or anticonformity. This normalization ensures that the probabilities define a complete behavioral strategy -- the agent must choose exactly one mode of behavior at each time step. Although, in principle, these probabilities could be defined independently and need not sum to one, such normalization provides an interpretable division of behavioral tendencies and simplifies analytical treatment of the model, allowing $p^{\mathrm{ind}}$, $p^{\mathrm{con}}$, and $p^{\mathrm{ant}}$ to be regarded as the expected fractions of updates governed by the respective mechanisms in the population. The parameter $q$ thus quantifies the size of the influence group and determines the strength of social pressure. 

An elementary update consists of the following consecutive steps:
\begin{enumerate}
    \item Choose one agent, denoted as $x$, by randomly drawing an integer from a uniform distribution in the range $\{1,\,\ldots,\,N\}$, expressed as $x \sim \mathcal{U}\{1$,~$N\}$.
    \item To determine whether an agent will act as a conformist, anticonformist or independently in the given update, draw a real random number, denoted as $r_1$, from a uniform distribution in the range $[0,\,1]$, i.e., $r_1 \sim \mathcal{U}(0$,~$1)$.
    \item If $r_1 < p^\mathrm{ind}$, the agent acts independently, as follows: 
    \begin{enumerate}
        \item Draw another real random number, denoted as $r_2$, from a uniform distribution in the range $[0,\,1]$, i.e.,  $r_2 \sim \mathcal{U}(0$,~$1)$.
        \item If $r_2 < p^\mathrm{eng}$, the agent adopts a state of $+1$; otherwise, it adopts a state of $-1$.
    \end{enumerate}
    \item Otherwise the agent can be influenced by others. If $q \le k_x$, choose $q$ neighbors out of the $k_x$ direct neighbors without repetitions.
    \item  If all $q$ chosen neighbors are in the same state, and
    \begin{enumerate}
        \item if $p^\mathrm{ind} \leq r_1 < p^\mathrm{ind} + p^\mathrm{ant}$, the agent will be anticonformist and take the opposite state to the group.
        \item if $p^\mathrm{ind} + p^\mathrm{ant} \leq r_1 < 1$, the agent will be conformist and take the state of the neighbors.
    \end{enumerate}
    \item Otherwise, when above options didn't occur, the agent remains uninfluenced and retains its previous state.
\end{enumerate}

A time period of a single update $\Delta t=1/N$, and one Monte Carlo step (MCS) is completed after $N$ elementary updates, i.e., when $t$ takes an integer value. After each MCS, data is collected, including information such as the concentration of adopted agents.


\section{Methods}
\label{sec:methods}
In this preprint, we examine the model using an approximated analytical approach, specifically the so-called mean-field approximation (MFA). This method has been widely applied in the study of complex systems, particularly for studying opinion or belief dynamics within agent-based models. A review of this method can be found in \cite{Jed:Szn:19}. The MFA serves as a useful starting point for observing the model’s behavior and comparing the results with the previous version without anticonformity \cite{abr-szu:wer:arxiv:25}, in order to identify potential differences. Its main advantage over Monte Carlo simulations is that analytical methods provide results much faster and do not require extensive computational resources.

\subsection{Analytical approach -- mean-field approximation}
\label{ssec:mfa}
Analytical methods allow to reduce the ABM to the mathematical model. MFA has been used to analyze the model without anticonformity in \cite{Byr:etal:16} and \cite{abr-szu:wer:arxiv:25}, but not to the model discussed in this paper. Therefore, here we will derive all formulas within the MFA formalism.

Within the MFA, the time evolution of the model can be expressed by one differential equation describing the time evolution of the concentration of adopted agents $c$, i.e., agents in state $+1$. In an elementary time step, the concentration 
$c$ may decrease or increase by $\Delta_c = 1/N$ or remain unchanged with respective probabilities derived for the infinite complete graph $N \to \infty$:
\begin{align}
\label{eq:gammas}
    \gamma^+ &= p^\mathrm{con} (1 - c) c^q + p^\mathrm{ant} (1-c)^{q+1} + p^\mathrm{ind} (1-c) p^\mathrm{eng}, \nonumber \\
    \gamma^- &= p^\mathrm{con} c (1-c)^q + p^\mathrm{ant} c^{q+1} + p^\mathrm{ind} c (1 - p^\mathrm{eng}), \\
    \gamma^0 &= 1 - \gamma^+ - \gamma^-. \nonumber
\end{align}
Using the transition probabilities defined above, the time evolution of the concentration of adopted agents can be expressed by the following rate equation
\begin{align}
\label{eq:general_mfa}
    \frac{d c}{d t} &= \gamma^+ - \gamma^- \\ \nonumber
    &= p^\mathrm{con} \left[ (1-c) c^q - c (1-c)^q \right] + p^\mathrm{ant} \left[ (1-c)^{q+1} - c^{q+1} \right] + p^\mathrm{ind} (p^\mathrm{eng} - c) \\ \nonumber
    &= (1  - p^\mathrm{ind} - p^\mathrm{ant}) \left[ (1-c) c^q - c (1-c)^q \right] + p^\mathrm{ant} \left[ (1-c)^{q+1} - c^{q+1} \right] + p^\mathrm{ind} (p^\mathrm{eng} - c).
\end{align}
Here, we have used the assumption that the probabilities sum to one, i.e., $p^{\mathrm{ind}} + p^{\mathrm{con}} + p^{\mathrm{ant}} = 1$.

Equation (\ref{eq:general_mfa}) not only describes the time evolution of the system but also allows us to determine its stationary states, which are obtained by the condition
\begin{equation}
\label{eq:stationary}
    \frac{dc}{dt} = 0.
\end{equation}
It is possible to derive the explicit formulas for $p^\mathrm{ind}(c)$ and $p^\mathrm{ant}(c)$ within MFA. We obtain:
\begin{align}
\label{eq:stacj:mfa:pind}
    \vphantom{\Bigl[} p^\mathrm{ind} &= \frac{(1 - p^\mathrm{ant}) \left[ (1-c) c^q - c (1-c)^q \right] + p^\mathrm{ant} \left[ (1-c)^{q+1} - c^{q+1} \right]}{(c - p^\mathrm{eng}) + \left[ (1-c) c^q - c (1-c)^q \right]},
\end{align}
\begin{align}
\label{eq:stacj:mfa:pant}
    \vphantom{\Bigl[} p^\mathrm{ant} &= \frac{(p^\mathrm{ind} - 1) \left[ (1-c) c^q - c (1-c)^q \right] + p^\mathrm{ind} (c - p^\mathrm{eng})}{\left[ (1-c)^{q+1} - c^{q+1} \right] - \left[ (1-c) c^q - c (1-c)^q \right]}.
\end{align}

If we put $p^\mathrm{ant} = 0$ into Eq. \eqref{eq:stacj:mfa:pind} we obtain the same formula for $p^\mathrm{ind}(c)$ as for the model without anticonformity \cite{abr-szu:wer:arxiv:25}, and if additionally we put $p^\mathrm{eng} = 1/2$, we obtain the same formula as for the model with symmetric independence \cite{Jed:17}.

\section{Results}
\label{sec:Results}
The analytical results derived above can be used to express the stationary concentration of adopters $c$ as a function of the model parameters.

\subsection{Trajectories}
For practitioners, the most interesting aspect is how the number of adopters changes over time, i.e., the trajectory $c(t)$. We need to verify if the trajectories obtained within our model exhibit a typical pattern that is observed in the real-life diffusion of innovation, namely the $S$-shaped curve of the number of adopters in time \cite{Rog:03}. 

\begin{figure*}[!ht]
    \centering
    \includegraphics[width=1\textwidth]{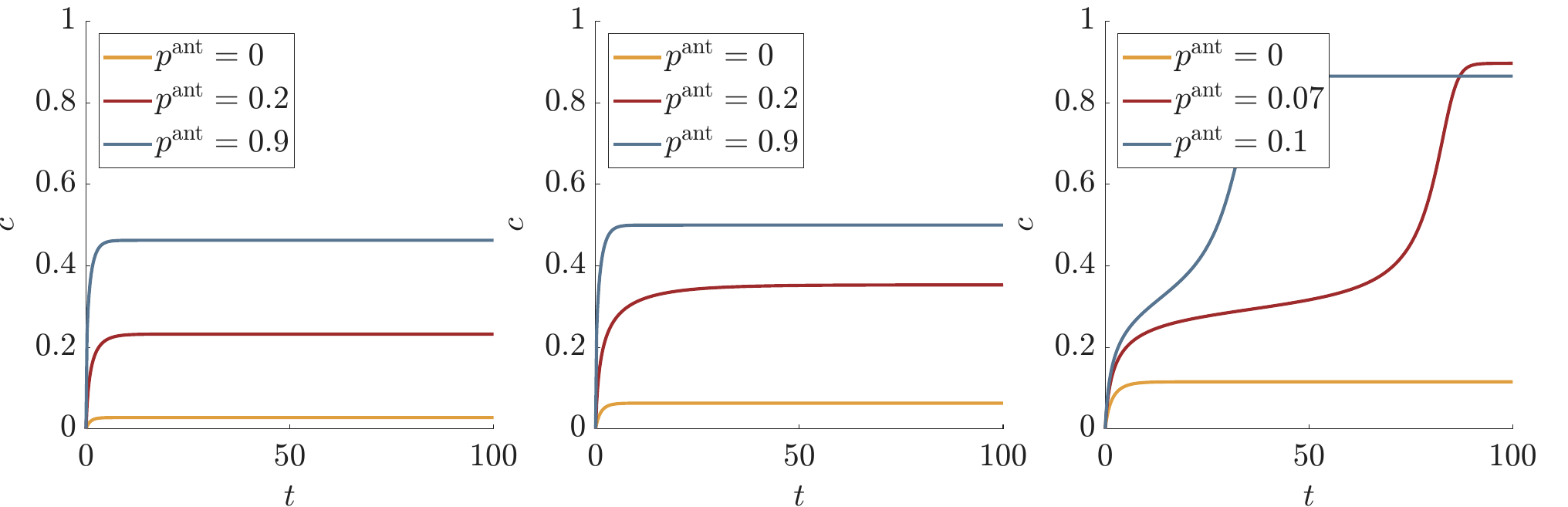}
    \caption{Results on the complete graph for the size of the influence group $q = 4$, the independence parameter $p^\mathrm{ind} = 0.1$, and three values of the engagement probability $p^\mathrm{eng} = 0.25$, $0.5$, $0.75$, respectively. Analytical average trajectories derived from MFA are shown for different values of anticonformity parameter $p^\mathrm{ant}$ -- in all cases the results are compared with the scenario for $p^\mathrm{ant} = 0$, corresponding to the model introduced in \cite{abr-szu:wer:arxiv:25}.} 
    \label{fig:Trajectories_mfa}
\end{figure*}

Trajectories can be obtained within the analytical approach, as shown in the Fig. \ref{fig:Trajectories_mfa}. We see that the number of adopted saturates at a certain level, depending on the parameters of the model. In some cases, it indeed produces the expected S-shaped curve of adoption over time. For 
$p^\mathrm{eng} = 0.75$ in the right panel of Fig. \ref{fig:Trajectories_mfa}, the results more closely resemble the patterns typically expected in the diffusion of innovation. Moreover, in each panel the final state depends on the level of anticonformity when all other initial conditions remain fixed. However, this dependence does not hold for all parameter values: for some combinations, such as e.g. $p^\mathrm{eng} = 0.5$ and selected values of $p^\mathrm{ind}$, the stationary level of $c$ remains the same regardless of $p^\mathrm{ant}$. Compared to the case without anticonformity ($p^\mathrm{ant} = 0$), increasing $p^\mathrm{ant}$ generally leads to a higher level of adoption. This result is rather surprising, as one might expect anticonformists to hinder consensus formation or diffusion \cite{Gal:23}. Instead, they initially accelerate the process and, in some cases, even enable successful diffusion in scenarios where no diffusion occurs for $p^\mathrm{ant} = 0$, with adoption stopping at a low level (see the right panel of Fig. \ref{fig:Trajectories_mfa}). As $p^\mathrm{ant}$ increases, diffusion becomes faster and more likely to succeed; however, its final level becomes slightly lower because anticonformists slow down adoption in the later phase. These findings highlight an important practical implication: decision-makers and policy designers may deliberately use a moderate level of anticonformity to accelerate early diffusion -- even at the cost of a slightly lower eventual adoption level -- depending on whether speed or maximal coverage is the priority in a given intervention. These findings are consistent with previous findings on the depolarizing effect of anticonformity \cite{Lip:Szn:22, Lip:Szn:25}.

\subsection{Phase diagram}
\begin{figure*}[!ht]
    \centering
    \includegraphics[width=1\textwidth]{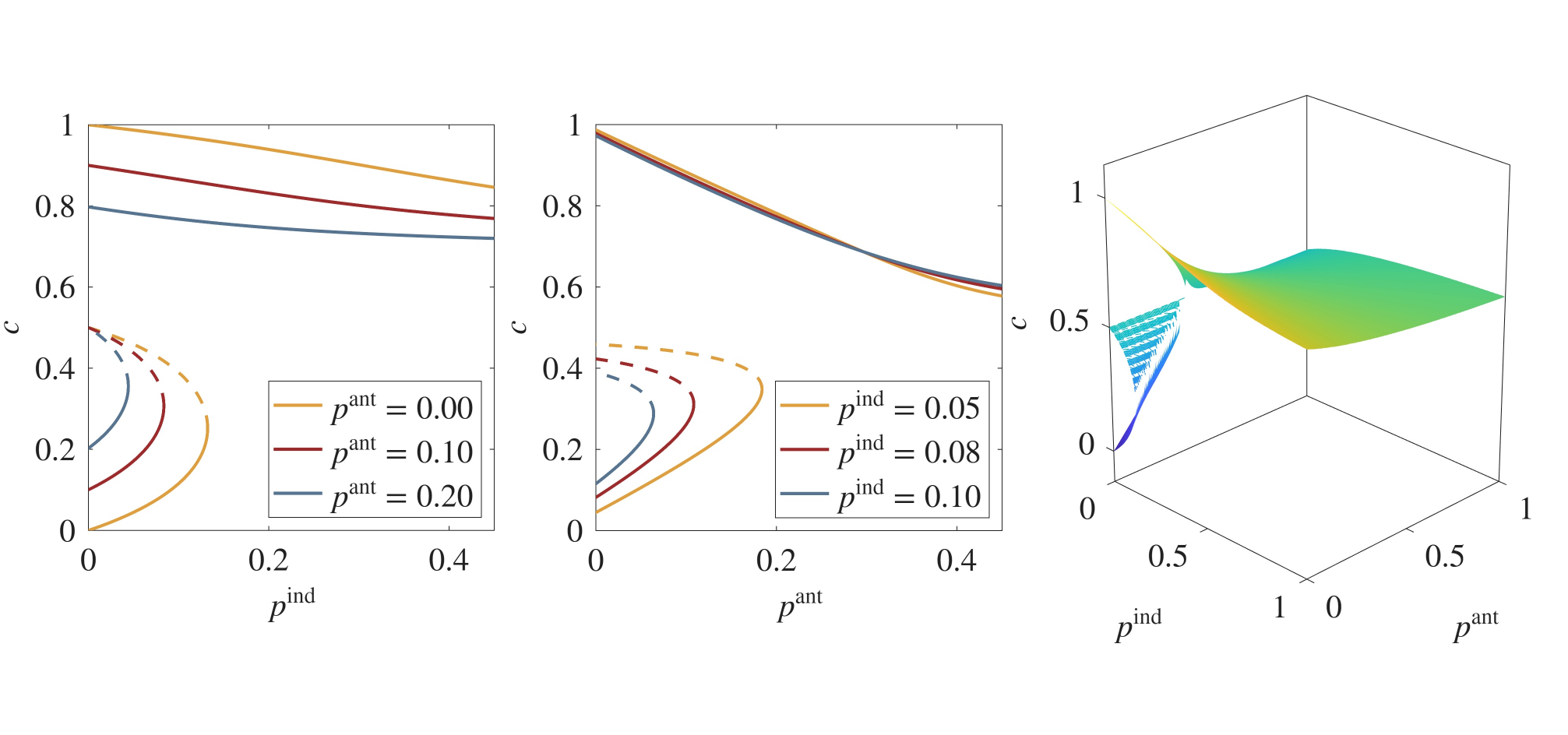}
    \caption{Phase diagrams on the complete graph for the influence group size $q = 4$ and the engagement probability $p^\mathrm{eng} = 0.75$, obtained from MFA. 
    \textbf{Left panel:} stationary fraction of adopters $c$ as a function of the independence parameter $p^\mathrm{ind}$, for three values of the anticonformity parameter $p^\mathrm{ant} = 0.0,\,0.1,\,0.2$. 
    \textbf{Middle panel:} stationary fraction of adopters $c$ as a function of the anticonformity parameter $p^\mathrm{ant}$, for three values of the independence parameter $p^\mathrm{ind} = 0.05,\,0.08,\,0.1$. 
    \textbf{Right panel:} stationary fraction of adopters $c$ as a function of both parameters, $c(p^\mathrm{ind},\,p^\mathrm{ant})$, across the full parameter range.}
    \label{fig:Stationary_mfa}
\end{figure*}

When analyzing the theoretical model systematically and conducting rigorous sensitivity analysis, we need a way to summarize the influence of the model's parameters on the system's behavior. One approach to achieve this is to examine the relationship between the stationary values of a macroscopic quantity, such as the concentration of adopters, and the model's parameters. This information is presented in Fig. \ref{fig:Stationary_mfa}. This type of a plot is often referred to as a phase diagram because it illustrates the system's phase, such as adopted or unadopted, for various parameter values. The lines in the plot represent solutions of Eq. (\ref{eq:stationary}), with solid lines indicating stable fixed points and dashed lines indicating unstable fixed points \cite{Str:94}.

In the left panel of Fig. \ref{fig:Stationary_mfa}, we observe that the function $c(p^\mathrm{ind})$ forms two branches of stable fixed points. The upper branch, denoted as $c_\mathrm{up}(p^\mathrm{ind})$, shows a slight decrease in $c$ as $p^\mathrm{ind}$ increases. The lower branch, $c_\mathrm{low}(p^\mathrm{ind})$, appears only for $p^\mathrm{ind} < p^\mathrm{ind}_\ast$. This means that there exists a critical value $p^\mathrm{ind} = p^\mathrm{ind}_\ast$, below which diffusion fails when the system is initialized with $c(0) = 0$, that is, when there are no initial adopters. In this regime, successful diffusion requires exceeding a critical mass of initial adopters -- the minimum initial number of adopters needed for diffusion to occur \cite{Mar:87}. Moreover, while the critical mass required increases with $p^\mathrm{ant}$, the interval $[0,\,p^\mathrm{ind}_\ast]$ in which such a critical mass is necessary becomes narrower as $p^\mathrm{ant}$ grows. Increasing the share of anticonformists therefore reduces the critical value $p^\mathrm{ind}_\ast$ compared to the case with $p^\mathrm{ant} = 0$. As a result, the system reaches the upper branch more easily and diffusion becomes successful for a wider range of $(p^\mathrm{ind},\,p^\mathrm{ant})$ combinations.

For $p^\mathrm{ind} < p^\mathrm{ind}_\ast$, the final state depends on the initial state. This implies that the system exhibits a kind of a collective memory, referred to as hysteresis \cite{Szn:Kam:Jed:23}. The unstable (repulsive) fixed points $c_\mathrm{rep}(p^\mathrm{ind})$ visible for $p^\mathrm{ind} < p^\mathrm{ind}_\ast$ and marked with dashed lines, indicate the existence of a critical mass. Such a threshold does not exist when $p^\mathrm{ind} > p^\mathrm{ind}_\ast$. If the system starts from an initial concentration $c_0 < c_\mathrm{rep}(p^\mathrm{ind})$, it converges to the lower branch $c_\mathrm{low}(p^\mathrm{ind})$, as illustrated in the left panel of Fig. \ref{fig:Stationary_mfa}. In contrast, if $c_0 > c_\mathrm{rep}(p^\mathrm{ind})$, the system moves toward the upper branch $c_\mathrm{up}(p^\mathrm{ind})$. Only when $c_0 = c_\mathrm{rep}(p^\mathrm{ind})$ does the system remain at this unstable point.

In the middle panel of Fig. \ref{fig:Stationary_mfa}, we show the dependence $c(p^\mathrm{ant})$ for several fixed values of $p^\mathrm{ind}$. The structure of this relationship is analogous to that in the left panel: two stable branches are present, with the lower one $c_\mathrm{low}(p^\mathrm{ant})$ existing only for $p^\mathrm{ant} < p^\mathrm{ant}_\ast$, and an unstable branch (dashed) separating their basins of attraction. Thus, a critical value $p^\mathrm{ant}_\ast$ emerges, below which the final state depends on the initial concentration of adopters, indicating hysteresis. Importantly, increasing $p^\mathrm{ind}$ shifts this threshold to lower values of $p^\mathrm{ant}$, meaning that independence facilitates diffusion: a smaller fraction of anticonformists is then sufficient for the system to reach the upper branch.

Finally, the right panel of Fig. \ref{fig:Stationary_mfa} shows the full surface $c(p^\mathrm{ind}, p^\mathrm{ant})$, combining the dependencies from the left and middle panels. The same structure is observed: two stable branches separated by an unstable branch (dashed), and the range of parameters where both branches exist becomes smaller as either $p^\mathrm{ind}$ or $p^\mathrm{ant}$ increases, confirming the consistent interplay between independence and anticonformity in shaping diffusion outcomes.

\clearpage

\section{Conclusion}
\label{sec:Conclusions}
In this study, we examine the role of anticonformist behavior in the diffusion of innovations by extending the $q$-voter model with asymmetric independence to allow agents not only to follow the group or act independently, but also to deliberately oppose peer pressure. This extension enabled us to systematically explore how different behavioral tendencies shape collective adoption dynamics in a social system.

Using the mean-field approximation, we showed that anticonformity -- typically viewed as a disruptive or noise-inducing force -- can, in fact, play a constructive role in diffusion. Consistent with recent findings in opinion dynamics, where anticonformity has been shown to depolarize rigidly divided groups or counteract social hysteresis, our results demonstrate that anticonformists can accelerate early adoption and enable successful diffusion even in scenarios where adoption would otherwise fail. In particular, anticonformity shrinks the region in which a critical mass is required for widespread adoption, thereby expanding the parameter space in which innovations can spread.

The model exhibits bistability and hysteresis, implying that the ultimate adoption level depends not only on behavioral tendencies but also on initial conditions. This highlights the importance of early interventions in promoting diffusion. More broadly, these findings show how social behaviors can either help or slow down collective change. The interplay between independence and anticonformity influences both the likelihood and the speed of diffusion, lowering the threshold of anticonformists needed to achieve widespread adoption.

From a practical perspective, our results suggest that moderate levels of anticonformist behavior can be beneficial for promoting innovation. Decision-makers and policymakers aiming to enhance the spread of new technologies or practices may therefore consider strategies that support diversity of opinion or tolerate the presence of anticonformists in the population. Future research could extend these findings to more realistic network structures or heterogeneous populations and validate the analytical results through Monte Carlo simulations to assess how well the mean-field approximation captures the actual model dynamics on various networks. Such extensions would provide a more comprehensive understanding of how behavioral tendencies interact with social structure to govern the spread of innovations.

\section*{Acknowledgments}
This work was supported by Polish Ministry of Science and Higher Education through project “Diamentowy Grant” no. DI2019 015049. 

\section*{Declaration of generative AI and AI-assisted technologies in the writing process}
During the preparation of this work the authors used DeepL Write and ChatGPT in order to improve the readability and language of certain parts of the work. After using these tools, the authors reviewed and edited the content as needed and take full responsibility for the content of the publication.

\bibliography{Szurlej_etal_bib}

\end{document}